\documentclass[traditabstract]{aa} 

\usepackage{graphicx, times, txfonts, float, rotating, color, url, lscape, bm}

\newcommand{\msol}{\mbox{M$_{\odot}$}}

\newcommand{\lsol}{\mbox{L$_{\odot}$}}

\newcommand{\G}{{\it Gaia}}

\begin{document} 
 
\title{The flux-weighted gravity-luminosity relation of Galactic classical Cepheids
\thanks{
Table~\ref{Tab:Sam} is available in electronic form at the CDS via 
anonymous ftp to cdsarc.u-strasbg.fr (130.79.128.5) or via 
http://cdsweb.u-strasbg.fr/cgi-bin/qcat?J/A+A/. 
Tables~\ref{Tab:fits} and \ref{Tab:FWGP}, and Figure~\ref{App:Fig} are available in the online edition of A\&A.
}
}  
 
\author{ 
M.~A.~T.~Groenewegen
}

\institute{ 
Koninklijke Sterrenwacht van Belgi\"e, Ringlaan 3, B--1180 Brussels, Belgium \\ \email{martin.groenewegen@oma.be}
} 
 
\date{received: ** 2020, accepted: * 2020} 
 
\offprints{Martin Groenewegen} 
 
 
\abstract
{
%
The flux-weighted gravity-luminosity relation (FWGLR) is investigated for a sample of 477 classical Cepheids (CCs), 
including stars that have been classified in the literature as such but are probably not.
The luminosities are taken from the literature, based on the fitting of the spectral energy distributions (SEDs)
assuming a certain distance and reddening. The flux-weighted gravity (FWG) is taken from gravity and effective temperature
determinations in the literature based on high-resolution spectroscopy.

There is a very good agreement between the theoretically predicted and observed FWG versus pulsation period relation that could serve
in estimating the FWG (and $\log g$) in spectroscopic studies with a precision of 0.1~dex.

As was known in the literature, the theoretically predicted FWGLR relation for CCs is very tight and is not very
sensitive to metallicity (at least for LMC and solar values), rotation rate, and crossing of the instability strip.
The observed relation has a slightly different slope and shows more scatter (0.54~dex). This is due both to uncertainties in the
distances and to the pulsation phase averaged FWG values.
Data from future {\it Gaia} data releases should reduce these errors, and then the FWGLR could serve
as a powerful tool in Cepheid studies.

}


\maketitle

\section{Introduction}
\label{S-Int}

Classical Cepheids (CCs) are considered an important standard candle because they are
bright, and thus they comprise a link between the distance scale in the nearby
universe and that further out via those galaxies that contain both Cepheids and SNIa 
(see  \citealt{Riess19} for a determination of the Hubble constant to 1.9\% precision, taking
into account the new 1.1\% precise distance to the Large Magellanic Cloud from \citealt{Pietrzynski19}).

This is the third paper in a series on Galactic CCs based on the {\it Gaia} second data
release (GDR2, \citealt{GDR2Sum}).
\citet{Gr_GDR2} (hereafter G18) started from an initial sample of 452 Galactic CCs with accurate [Fe/H] abundances from
spectroscopic analysis. Based on parallax data from \G\ DR2, supplemented with accurate non-\G\ parallax data when available,
a final sample of about 200 FU mode Cepheids with good astrometric solutions was retained to derive period-luminosity ($PL$) and
period-luminosity-metallicity ($PLZ$) relations. 
The influence of a parallax zeropoint offset on the derived $PL(Z)$ relation is large and means that the 
current GDR2 results  do not allow to improve on the existing calibration of the relation or on the distance to the LMC
(as also concluded by \citealt{RiessGDR2}). The zeropoint, the slope of the period dependence, and the metallicity dependence
of the $PL(Z)$ relations are correlated with any assumed parallax zeropoint offset.

In \citet{Gr20} (hereafter G20) the sample was expanded to 477 stars. Using photometry over the widest available
range in wavelength (and at mean light when available) the spectral energy distributions (SEDs) were constructed and
fitted with model atmospheres (and a dust component when required). For an adopted distance and reddening these fits resulted in a
best-fitting bolometric luminosity ($L$) and the photometrically derived effective temperature ($T_{\rm eff}$).
This allowed for the derivation of
period-radius ($PR$) and $PL$ relations, the construction of the Hertzsprung-Russell diagram (HRD), and a comparison to
theoretical instability strips (ISs). The position of most stars in the HRD was consistent with theoretical predictions.
Outliers were often associated with sources where the spectroscopically and photometrically  determined
effective temperatures differed, or with sources with large and uncertain reddenings.

In this paper the relation between bolometric absolute magnitude and the flux-weighted gravity (FWG),
$g_{\rm F} \sim g/T_{\rm eff}^4$, is investigated: the so-called flux-weighted gravity-luminosity relation (FWGLR).
The tight correlation between $g_{\rm F}$ and luminosity was first
demonstrated by \citet{K2003,K2008} for blue supergiants, and was then used for extragalactic distance determinations in \citet{K16}. 
\citet{Anderson16} demonstrated that theoretical pulsation models for CCs also followed a tight FWGLR, in fact tighter than
the $PL$ relation, and that there was a good correspondence between observed $g_{\rm F}$ and period for a sample of CCs.
The latest calibration of the FWGLR is presented in \citet{K20} based on 445 stars ranging from $M_{\rm bol}= +9.0$ to $-8.0$. 

The paper is structured as follows.
In Section~\ref{S-theory} the theoretical models of \citet{Anderson16} are compared to the latest calibration in \citet{K20}.
In Section~\ref{S-Sam} the sample of 477 (candidate) CCs is introduced and the $g_{\rm F}$ are derived, and the
correlation with period and luminosity are presented.
A brief discussion and summary concludes the paper.

\section{Theoretical FWGLR for CCs} 
\label{S-theory}

The FWG is defined as
$\log g_{\rm F} = \log g - 4 \cdot \log \left(T_{\rm eff} / 10^4\right)$ \citep{K2003}.
\citet{K20} present the latest calibration of the FWG against absolute bolometric magnitude as
\begin{equation}
  M_{\rm bol} =  (3.19 \pm 0.01) (\log g_{\rm F} - g_{\rm F}^{\rm \odot}) + (4.74 \pm 0.01) 
\end{equation}
for $\log  g_{\rm F} \ge \log  g_{\rm F}^{\rm b}$ and
\begin{equation}
  M_{\rm bol} =  (3.76 \pm 0.11) (\log g_{\rm F} - g_{\rm F}^{\rm b})    + (-2.98 \pm 0.09) 
\end{equation}
for $\log  g_{\rm F} <   \log  g_{\rm F}^{\rm b}$, with a scatter of 0.17 and 0.29~mag, respectively.
The break in the relation is set at $g_{\rm F}^{\rm b} = 3.0$, while the FWG of the Sun is $g_{\rm F}^{\rm \odot} = 5.39$.

\citet{Anderson16} presented a large set of pulsation models for CCs based on stellar evolutionary models for a range of initial
masses (1.7-15~\msol), initial rotation rates ($\omega_{\rm ini} = 0.0, 0.5, 0.9$ in terms of the critical rotation rates),
metallicities ($Z= 0.002, 0.006, 0.014$), and for fundamental mode (FU) and first overtone (FO) CCs.
Stellar parameters ($L$, $T_{\rm eff}$), and pulsation periods are given at the entry and exit of the IS for various crossings.
They used these models to show the tight FWGLR for CCs for the first time (Fig.~16 in \citealt{Anderson16}).

\begin{figure}
  \centering

\begin{minipage}{0.49\textwidth}
\resizebox{\hsize}{!}{\includegraphics{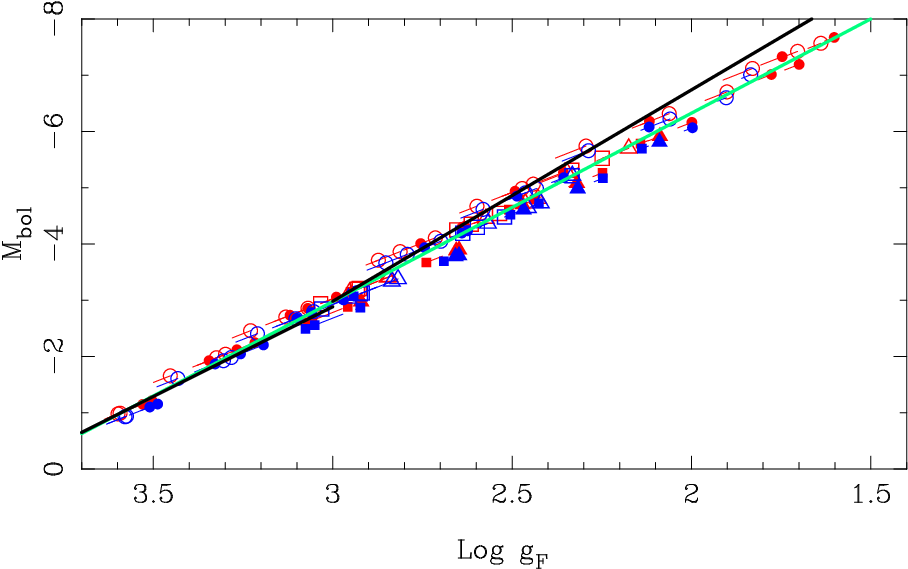}}
\resizebox{\hsize}{!}{\includegraphics{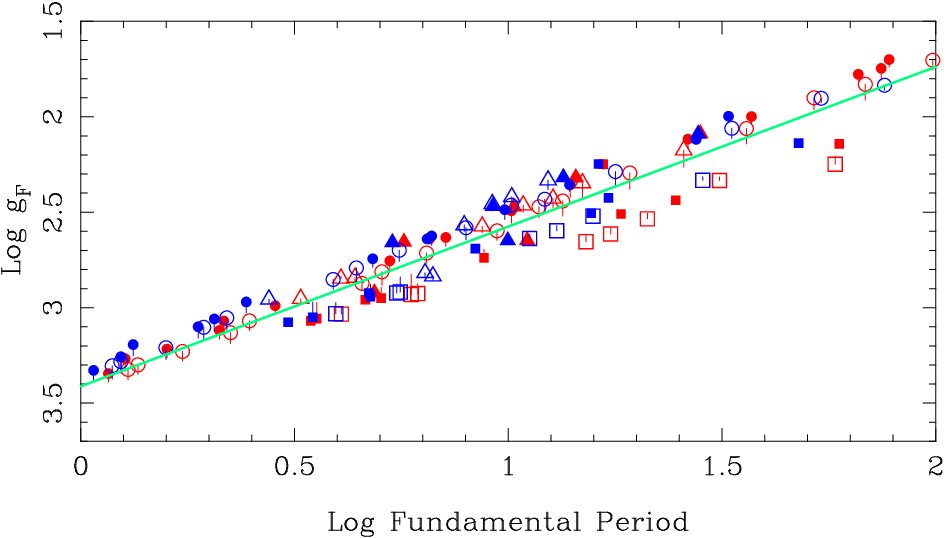}}
\end{minipage}

\caption{{\it Top panel.} FWGLR based on the pulsation models in \citet{Anderson16}.
  FU models are shown in red, FO models are shown in blue.
  For clarity FU (FO) models are plotted with an offset of +0.05 ($-0.05$)~dex in $M_{\rm bol}$.
  Symbols indicate the entry point of the IS, the lines connect it to the exit point of the IS.
  The first, second, and third crossing models are plotted as circles, squares, and triangles, respectively.
  Solar metallicity models are plotted with open symbols, models with $Z=0.006$ with filled symbols.
  The black lines refer to Eqs.~1 and 2, the green line to the best fit (Eq.~\ref{Eq:And}).
  {\it Bottom panel.} Relation between FWG and period for the same models. The period of the FO models was fundamentalised.
   The green line refers to the best fit, Eq.~\ref{Eq:FWGPer}.
}
\label{Fig:FGWLR}
\end{figure}

The top panel in Fig.~\ref{Fig:FGWLR} shows the theoretical FWGLR based on these models for FU pulsators with periods $>0.6$~d,
FO pulsators with periods $>0.4$~d, $Z= 0.006$ and $0.014$, and all rotation rates and crossings of the IS as the coloured
lines and symbols. Also shown are Eqs.~1 and 2.
For the lower gravities the models deviate from Eq.~2, and appear to be closer to an extension of Eq.~1.
A linear fit to these models gives the relation
\begin{equation}
  M_{\rm bol} =  (3.35 \pm 0.02) (\log g_{\rm F} - g_{\rm F}^{\rm b})    + (-2.975 \pm 0.012)
\label{Eq:And}
\end{equation}
with an rms of 0.16~mag, shown as the green line in the figure.
Additional fits are given in Appendix~\ref{App}.

The bottom panel shows the relation between FWG and period for the same selection of models (cf. Figure~17 in \citet{Anderson16,Anderson20}).
Periods of FO models are fundamentalised using the relation $P_0 = P_1/(0.716-0.027\, \log P_1)$ following \citet{FC1997}.
The best fit is
\begin{equation}
  \log g_{\rm F} =  (-0.834 \pm 0.011) \;\log P_{\rm 0} + (3.402  \pm 0.011)
\label{Eq:FWGPer}
\end{equation}
with an rms of 0.09~dex. 
Eliminating the second crossing models would result in a fit with a smaller scatter, but as this information is not known a priori the
relation as presented is more generally applicable when an estimate of $\log g_{\rm F}$ is desired.
Figures and relations for FU and FO models separately are presented in the Appendix.

\section{Sample and observed FWGs.} 
\label{S-Sam}

The sample studied here is the collection of 477 stars considered in G20. It is based on the original sample of 452 stars
compiled in G18, extended by 25 additional stars for which accurate iron abundances have since become available, including
five CCs in the inner disk of our Galaxy \citep{Inno19}.

G20 constructed the SEDs for these stars, considering photometry from the ultraviolet to the far-infrared,
and as much as possible at mean light. Distances and reddening were collected from the literature. Distances from GDR2
data was available for 232 sources, and from other parallax data for 26 stars.

\citet{Luck18} (hereafter L18) published a list of abundances and stellar parameters for 435 Cepheids based on the analysis
of 1137 spectra.
L18 reduced all data in a uniform way using MARCS LTE model atmospheres \citep{Gustafsson_MARCS}.
Effective temperatures were determined in that paper using the line depth ratio (LDR) -- effective temperature calibration
of \citet{Kov07} as updated by Kovtyukh (2010, private communication to Luck), while
gravities were determined from the ionisation balance between Fe I and Fe II  lines, and micro-turbulent velocities ($v_{\rm t}$)
by forcing there to be no dependence in the per-line Fe I abundances on equivalent width (see L18 for additional details).

\begin{figure}
  \centering

\begin{minipage}{0.42\textwidth}
\resizebox{\hsize}{!}{\includegraphics{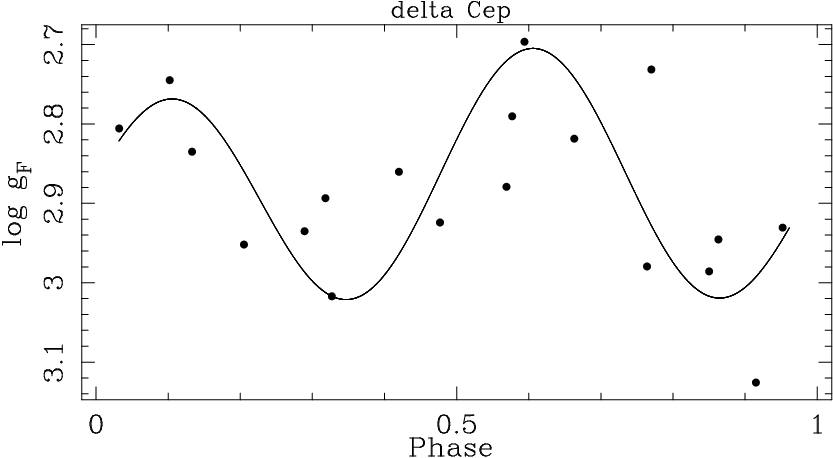}}
\resizebox{\hsize}{!}{\includegraphics{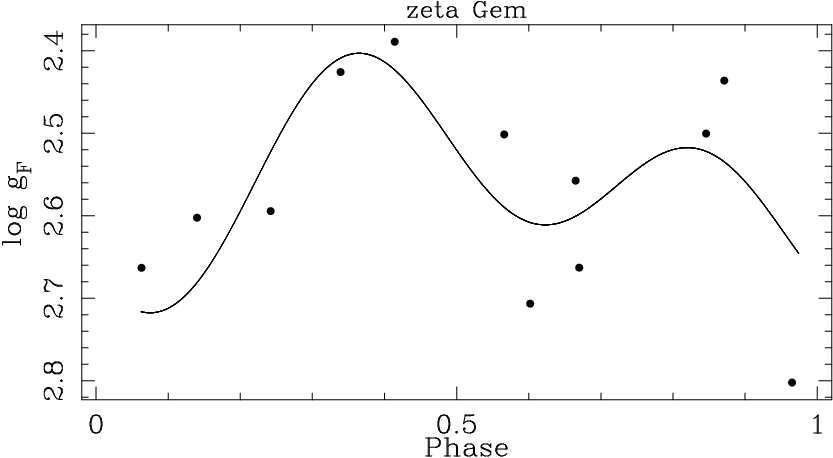}}
\resizebox{\hsize}{!}{\includegraphics{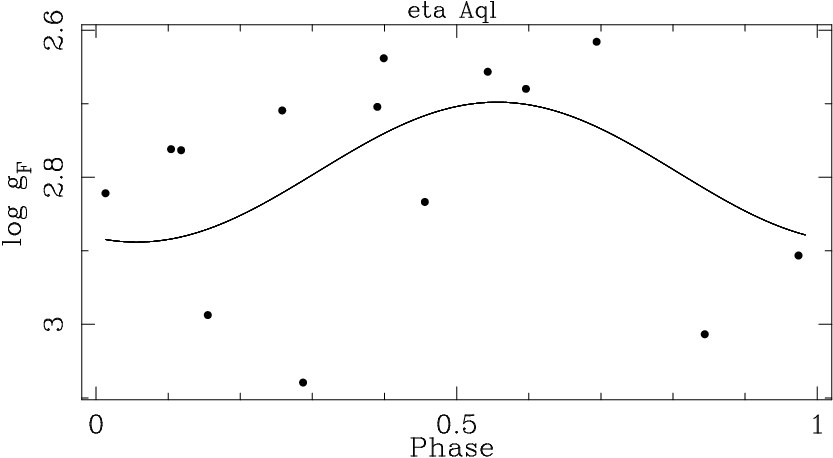}}
\resizebox{\hsize}{!}{\includegraphics{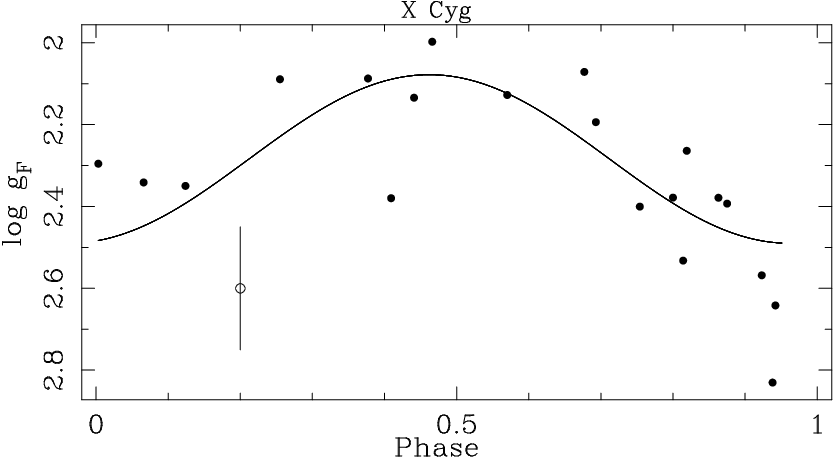}}
\end{minipage}

\caption{FWG vs pulsation phase for four CCs.
  The typical error bar in each point is 0.15~dex in FWG, as indicated in the bottom plot.
    The lines are low-order harmonic fits to the data (see Col.~7 in Table~\ref{Tab:FGLR}). 
}
\label{Fig:loggf}
\end{figure}

Table~\ref{Tab:FGLR} contains information on the set of 52 CCs for which five or more spectra were available in L18 
taken at different phases in the pulsation cycle. 
FWGs are calculated on the one hand from the mean effective temperatures
and mean gravities (as given by L18 in his Table~11), and on the other hand from an analysis of the FWGs calculated for
the individual epochs and plotted versus phase.
Using the code {\sc Period04} \citep{Period04} to fit a low-order harmonic, this gives the mean $\log g_{\rm F}$, the amplitude of
the $\log g_{\rm F}$ curve, and the rms value. Some $\log g_{\rm F}$ phased curves with fits are shown in Fig.~\ref{Fig:loggf}.

\begin{table*}
  \centering

    \small
\setlength{\tabcolsep}{1.6mm}
\caption{FWGL data for the subsample with more than five spectra.}
\begin{tabular}{rrrcccccccr} \hline \hline 
  Name   & Period & $N_{\rm spec}$   & $<T_{\rm eff}>$ & $<\log g>$ & $<\log g_{\rm F}>$ & $N_{\rm h}$ & $\log g_{\rm F}$ & Ampl & rms & Luminosity  \\
         &  (days) &               &     (K)       &  (cgs)     &      (cgs)        &           &      (cgs)    &   (cgs) & (cgs) & (\lsol) \\
  \hline                                                                                          
V473 Lyr  & 1.490780 &  5 & 6019 & 2.30 & 3.18 & 1 & 3.205 & 0.056 & 0.006 & 572.3 $\pm$ 10.8 \\
SU Cas    & 1.949324 & 13 & 6274 & 2.26 & 3.07 & 2 & 3.065 & 0.122 & 0.046 & 1027.0 $\pm$ 16.6 \\ 
DT Cyg    & 2.499215 & 14 & 6192 & 2.27 & 3.10 & 1 & 3.108 & 0.049 & 0.066 & 157.6 $\pm$ 70.7 \\  
SZ Tau    & 3.148730 & 16 & 5987 & 2.03 & 2.92 & 1 & 2.937 & 0.079 & 0.081 & 1155.3 $\pm$ 23.1 \\ 
V1334 Cyg & 3.332816 & 11 & 6293 & 2.22 & 3.02 & 1 & 3.014 & 0.119 & 0.082 & 1879.5 $\pm$ 384.2 \\ 
RT Aur    & 3.728485 & 12 & 5948 & 2.06 & 2.96 & 1 & 2.978 & 0.048 & 0.031 & 970.8 $\pm$ 46.1 \\  
SU Cyg & 3.845547 & 12 & 6036 & 2.08 & 2.96 & 1 & 2.988 & 0.067 & 0.136 & 909.6 $\pm$ 294.7 \\
ST Tau & 4.034299 & 7 & 6052 & 2.07 & 2.94 & 1 & 2.910 & 0.078 & 0.059 & 1155.3 $\pm$ 23.1 \\
BQ Ser & 4.270900 & 7 & 6040 & 2.16 & 3.04 & 1 & 2.860 & 0.294 & 0.129 & 1876.3 $\pm$ 97.1 \\
Y Lac  & 4.323776 & 10 & 5915 & 1.87 & 2.78 & 1 & 2.779 & 0.111 & 0.108 & 1250.6 $\pm$ 82.7 \\
T Vul  & 4.435462 & 12 & 5852 & 2.03 & 2.96 & 1 & 2.949 & 0.131 & 0.109 & 691.9 $\pm$ 27.6 \\     
FF Aql & 4.470881 & 14 & 6164 & 2.04 & 2.88 & 1 & 2.886 & 0.085 & 0.086 & 2237.6 $\pm$ 118.3 \\   
CF Cas & 4.875220 & 7 & 5672 & 1.74 & 2.73 & 1 & 2.748 & 0.129 & 0.142 & 1284.7 $\pm$ 41.1 \\
BG Lac & 5.331908 & 9 & 5674 & 1.70 & 2.68 & 2 & 2.710 & 0.218 & 0.035 & 1303.6 $\pm$ 43.2 \\
delta Cep & 5.366341 & 19 & 5854 & 1.96 & 2.89 & 3 & 2.878 & 0.108 & 0.068 & 1984.8 $\pm$ 587.1 \\ 
Y Sgr  & 5.773350 & 12 & 5767 & 1.77 & 2.73 & 1 & 2.724 & 0.088 & 0.096 & 1737.4 $\pm$ 90.1 \\
FM Aql & 6.114290 & 12 & 5766 & 1.68 & 2.64 & 1 & 2.667 & 0.085 & 0.128 & 2575.5 $\pm$ 91.8 \\
X Vul  & 6.319588 & 8 & 5753 & 1.81 & 2.77 & 1 & 2.774 & 0.074 & 0.113 & 1735.3 $\pm$ 81.7 \\
XX Sgr & 6.424140 & 5 & 5805 & 1.81 & 2.75 & 1 & 2.671 & 0.126 & 0.032 & 1913.5 $\pm$ 123.3 \\
AW Per & 6.463589 & 11 & 5928 & 1.86 & 2.77 & 2 & 2.776 & 0.153 & 0.079 & 1646.8 $\pm$ 76.6 \\
U Sgr  & 6.745226 & 11 & 5709 & 1.79 & 2.76 & 2 & 2.769 & 0.178 & 0.086 & 2421.8 $\pm$ 69.6 \\
U Aql  & 7.024049 & 5 & 5565 & 1.64 & 2.66 & 1 & 2.654 & 0.077 & 0.026 & 1619.1 $\pm$ 42.2 \\
eta Aql & 7.176915 & 14 & 5746 & 1.86 & 2.82 & 1 & 2.793 & 0.095 & 0.124 & 3008.4 $\pm$ 598.2 \\  
BB Her  & 7.507945 & 8 & 5641 & 1.65 & 2.64 & 1 & 2.655 & 0.075 & 0.091 & 3122.0 $\pm$ 153.2 \\
RS Ori  & 7.566881 & 7 & 5891 & 1.77 & 2.69 & 1 & 2.693 & 0.109 & 0.112 & 2683.4 $\pm$ 85.8 \\
V440 Per & 7.570000 & 10 & 6056 & 1.97 & 2.84 & 1 & 2.829 & 0.087 & 0.109 & 3257.4 $\pm$ 50.3 \\
W Sgr   & 7.595030 & 9 & 5765 & 1.78 & 2.74 & 1 & 2.726 & 0.096 & 0.126 & 3277.6 $\pm$ 312.9 \\
RX Cam  & 7.912024 & 10 & 5703 & 1.65 & 2.63 & 1 & 2.642 & 0.166 & 0.087 & 2192.7 $\pm$ 87.0 \\
W Gem   & 7.913779 & 13 & 5771 & 1.69 & 2.64 & 1 & 2.662 & 0.131 & 0.074 & 3631.9 $\pm$ 179.0 \\
U Vul   & 7.990676 & 8 & 5779 & 1.73 & 2.68 & 1 & 2.673 & 0.042 & 0.069 & 5408.2 $\pm$ 241.2 \\
DL Cas  & 8.000669 & 11 & 5682 & 1.56 & 2.54 & 2 & 2.566 & 0.189 & 0.049 & 1957.5 $\pm$ 49.4 \\
V636 Cas & 8.375710 & 8 & 5505 & 1.47 & 2.51 & 1 & 2.508 & 0.036 & 0.038 & 3268.0 $\pm$ 81.6 \\
S Sge   & 8.382086 & 11 & 5689 & 1.73 & 2.71 & 1 & 2.743 & 0.133 & 0.098 & 2286.1 $\pm$ 57.8 \\   
V500 Sco & 9.316863 & 5 & 5675 & 1.56 & 2.54 & 1 & 2.543 & 0.042 & 0.062 & 2654.7 $\pm$ 143.3 \\
FN Aql   & 9.481640 & 11 & 5488 & 1.38 & 2.42 & 1 & 2.456 & 0.181 & 0.106 & 2699.4 $\pm$ 68.5 \\
YZ Sgr   & 9.553606 & 11 & 5653 & 1.69 & 2.68 & 1 & 2.681 & 0.026 & 0.074 & 3438.5 $\pm$ 126.8 \\
zeta Gem & 10.150730 & 12 & 5512 & 1.52 & 2.55 & 2 & 2.562 & 0.126 & 0.080 & 3203.4 $\pm$ 928.2 \\ 
Z Lac  & 10.885613 & 10 & 5618 & 1.49 & 2.49 & 2 & 2.499 & 0.119 & 0.058 & 4173.7 $\pm$ 193.5 \\
VX Per & 10.889040 & 12 & 5783 & 1.64 & 2.59 & 1 & 2.579 & 0.130 & 0.147 & 4407.1 $\pm$ 107.4 \\
RX Aur & 11.626000 & 13 & 5782 & 1.67 & 2.62 & 1 & 2.623 & 0.155 & 0.085 & 4670.0 $\pm$ 204.7 \\
TT Aql & 13.754912 & 10 & 5272 & 1.15 & 2.26 & 2 & 2.400 & 0.402 & 0.104 & 5242.1 $\pm$ 206.0 \\
SV Mon & 15.232780 &  9 & 5330 & 1.11 & 2.20 & 1 & 2.220 & 0.237 & 0.136 & 4952.3 $\pm$ 352.7 \\
X Cyg  & 16.386332 & 20 & 5252 & 1.10 & 2.22 & 1 & 2.284 & 0.206 & 0.140 & 5201.9 $\pm$ 280.9 \\  
RW Cam & 16.415014 & 17 & 5213 & 1.03 & 2.16 & 1 & 2.200 & 0.100 & 0.156 & 4857.7 $\pm$ 187.4 \\
CD Cyg & 17.073967 & 17 & 5394 & 1.19 & 2.26 & 2 & 2.230 & 0.270 & 0.103 & 5399.7 $\pm$ 191.6 \\
Y Oph  & 17.124130 & 14 & 5819 & 1.62 & 2.56 & 1 & 2.561 & 0.088 & 0.061 & 12857.9 $\pm$ 388.4 \\
SZ Aql & 17.141247 & 11 & 5398 & 1.20 & 2.27 & 2 & 2.299 & 0.150 & 0.084 & 7077.7 $\pm$ 232.3 \\
WZ Sgr & 21.849709 & 10 & 5140 & 0.88 & 2.04 & 2 & 2.204 & 0.514 & 0.060 & 8349.1 $\pm$ 239.9 \\
X Pup  & 25.961000 &  8 & 5353 & 0.75 & 1.84 & 1 & 1.923 & 0.374 & 0.073 & 9419.5 $\pm$ 552.9 \\
T Mon  & 27.024649 & 12 & 5108 & 0.93 & 2.10 & 1 & 2.115 & 0.162 & 0.141 & 8163.2 $\pm$ 203.3 \\
SV Vul & 45.012100 & 15 & 5329 & 0.85 & 1.94 & 1 & 1.905 & 0.124 & 0.157 & 27925.1 $\pm$ 1818.3 \\
S Vul  & 68.463997 &  6 & 5452 & 0.93 & 1.98 & 1 & 1.929 & 0.281 & 0.082 & 21197.2 $\pm$ 747.9 \\
\hline
\end{tabular} 
\tablefoot{
  Column~1: Name.
  Column~2: Period (as quoted in L18).
  Column~3: Number of spectra L18.
  Column~4: Average effective temperature  (quoted in Table~11 in L18).
  Column~5: Average $\log g$ (quoted in Table~11 in L18).
  Column~6: Average $\log g_{\rm F}$ based on Cols.~4 and 5.
  Column~7: Number of harmonics used in the time analysis.
  Column~8: Mean $\log g_{\rm F}$.
  Column~9: Amplitude in the $\log g_{\rm F}$ curve.
  Column~10: RMS.
  Column~11: Luminosity and error (from Table~1 in G20).
  The error is the fit error, and does not include the error on the distance.
  The distance and error on the distance needed to calculate the total error on $L$ are given in Table~\ref{Tab:Sam}. 
}
\label{Tab:FGLR}
\end{table*}

These curves show considerable scatter even when the pulsation cycle is well sampled. This is likely due to the error bar in an individual
determination of $g_{\rm F}$.
The error on effective temperature generally has a negligible contribution in this. Ninety-five percent of individual effective
temperature error bars among the 1137 spectra in L18 are between 30 and 220~K with a median of 65~K.
An error of 100~K at $T_{\rm eff}= 6000$~K introduces an error of 0.03~dex in $g_{\rm F}$, much smaller than
the error on $\log g$, which was estimated to be $\sim$0.15~dex by L18.
A comparison of $\log g_{\rm F}$ values determined from the averages of the effective temperatures and gravities, and from fitting the 
$\log g_{\rm F}$ curve with phase show essentially the same result, especially when seven of more spectra are averaged
 (with an average difference between Cols.~6 and 8 of $-0.02 \pm 0.04$~dex).

\begin{figure}
  \centering

\begin{minipage}{0.49\textwidth}
\resizebox{\hsize}{!}{\includegraphics{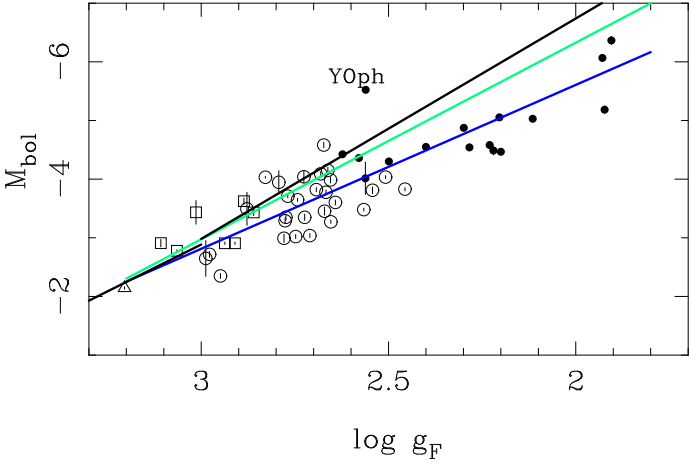}}
\end{minipage}

\caption{FWGLR based on the subsample with more than five spectra.
  FU mode pulsators are plotted as circles (filled circles for periods over 10 days),
  FO pulsators as open squares, the single second-overtone pulsator as open triangle.
  The black lines refer to Eqs.~1 and 2, the green line to Eq.~\ref{Eq:And}.
  The blue line is a fit to the data points, excluding Y Oph (Eq.~\ref{Eq:data}).
}
\label{Fig:FGLR}
\end{figure}

Figure~\ref{Fig:FGLR} shows the FWGLR for the sample of 52 stars from Table~\ref{Tab:FGLR}, where the luminosity and error
are taken from G20.
Equations~1, 2, and \ref{Eq:And} are plotted as reference.
Using a linear bi-sector fit (using the code SIXLIN from \citealt{Isobe90}) the best fit is
\begin{equation}
  M_{\rm bol} =  (2.79 \pm 0.18) (\log g_{\rm F} - 2.5)    + (-4.21 \pm 0.08)
\label{Eq:data}
\end{equation}
with an rms of 0.38~mag (blue line in the figure).
A standard least-squares fit has a shallower slope of $2.54$.
The theoretical fit is shown in Eq.~\ref{Eq:And}, and this fit differs by about 0.4~mag at $\log g_{\rm F}= 2.5$.
Alternatively, the observed $\log g_{\rm F}$ values are systematically too small by $0.4/2.8 \sim 0.14$~dex.
At lower FWG or longer periods the difference with the theoretical relation is larger.

\begin{figure}

\begin{minipage}{0.49\textwidth}
\resizebox{\hsize}{!}{\includegraphics{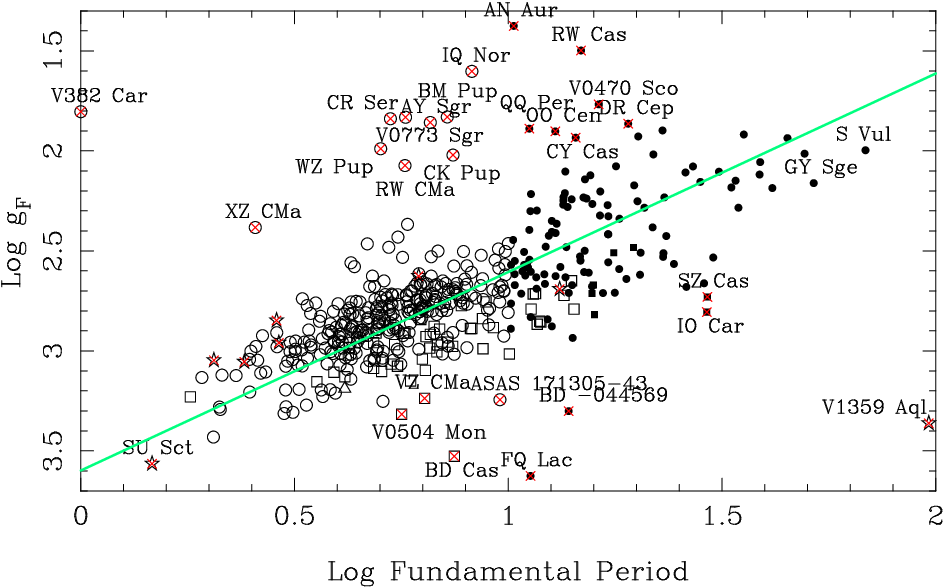}}
\end{minipage}

\caption{FWG vs fundamental pulsation period.
  Some outliers are named.
  The green line refers to the best fit, Eq.~\ref{Eq:FWGPerObs}, which excludes the outliers and non-CCs
  indicated by a red cross.
}
\label{Fig:loggfper}
\end{figure}

Table~\ref{Tab:Sam} collects the FWG data for the entire sample of 477 stars. Overall, most of the data (435 stars)
come from L18, and for the remaining stars $\log g$ and $T_{\rm eff}$ have been
collected from the literature in order to calculate $\log g_{\rm F}$.
Multiple determinations of $\log g_{\rm F}$  have been averaged and so can differ slightly from the values
in Table~\ref{Tab:FGLR}.
The table also includes the period, pulsation type, distance with error, and luminosity with error from G20. 
Figure~\ref{Fig:loggfper} shows the observational equivalent to the bottom panel in Fig.~\ref{Fig:FGWLR},
the FWG determined from spectroscopy against pulsation period (fundamentalised for FO pulsators).
  
There is a tight correlation between the two quantities. Removing non-CCs (see Table~\ref{Tab:Sam}) and applying iterative 3$\sigma$
clipping results in the fit
\begin{equation}
  \log g_{\rm F} =  (-0.80 \pm 0.03) \;\log P_{\rm 0} + (3.43  \pm 0.03)
\label{Eq:FWGPerObs}
\end{equation}
with an rms of 0.16~dex, in very good agreement with the theoretically predicted relation. 
Interestingly, many of the outliers come from a single source, \cite{Genovali2014}, who derived very low $\log g$ values for some objects.
Some additional information and fits are provided in Appendix~\ref{App}.

Figure~\ref{Fig:loggfobs} is the equivalent to Fig.~\ref{Fig:FGLR} for the entire sample, using
a simple averaging of the available FWGs. The error on distance is now taken into account in calculating the error
on luminosity. Following the discussion above and in the Appendix, the data from \cite{Genovali2014} has been excluded,
and to reduce the scatter only stars with two or more spectra are considered.
A linear bi-sector fit  applying iterative 3$\sigma$ clipping results in
\begin{equation}
  M_{\rm bol} =  (2.93 \pm 0.13) (\log g_{\rm F} - 2.5)    + (-4.23 \pm 0.06)  
\label{Eq:FWGPerObs}
\end{equation}
with an rms of 0.54~mag using 170 stars and is shown as the blue line in the figure.
This is currently the best observational determination of the FWGL relation for CCs.

\begin{figure}

\begin{minipage}{0.49\textwidth}
\resizebox{\hsize}{!}{\includegraphics{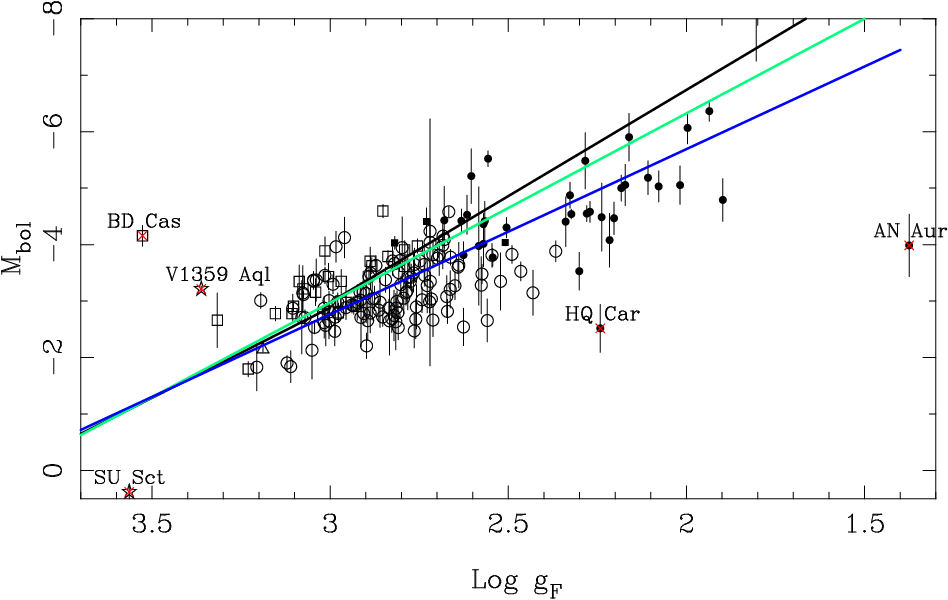}}
\end{minipage}

\caption{FWGLR, with some outliers named.
  The black lines refer to Eqs.~1 and 2, the green line to Eq.~\ref{Eq:And}.
  The blue line refers to the best fit, Eq.~\ref{Eq:FWGPerObs}, which excludes the outliers and non-CCs
  indicated by a red cross. 
  %
  Outliers located outside the plot window are
SU Cru   ($\log g_{\rm F}= 0.19$, $M_{\rm bol}= -3.7$),  
SY Nor   ($\log g_{\rm F}= 2.4$, $M_{\rm bol}= +3.3$), and
V382 Car ($\log g_{\rm F}= 1.8$, $M_{\rm bol}= -8.6$).
}
\label{Fig:loggfobs}
\end{figure}

\section{Discussion and summary}
\label{S-Dis}

The relation between FWG and period, and FWG and bolometric luminosity is investigated for a sample 477 CCs.
The FWGs are derived from effective temperatures and $\log g$ values available in the literature based on
high-resolution spectroscopy. The overall majority of parameters have been compiled from a single source (L18) that
  determined $\log g$ and $T_{\rm eff}$
in an uniform manner. For a subset of stars multiple-phase data is available.
The FWG-Period and FWGLR are compared to theoretical models from \citet{Anderson16}

A very good agreement is found between the theoretical and observed relations between FWG and period.
These relations could serve as a prediction for a reasonable range in $\log g$ values (assuming an effective
temperature) in a spectroscopic analysis.

The observed FWGLR is found to have a shallower slope than the theoretical relation.
It is not clear at the moment if this is a significant effect or not. As the observed relation between
FWG and period agrees with the theoretical relation, one would be inclined to think that there could be a
systematic effect in the bolometric magnitudes of the long-period Cepheids. They are rarer and on average
at longer distance, likely to be more susceptible to (systematic) errors on parallax.
This is qualitatively confirmed by repeating the fit of Eq.~\ref{Eq:FWGPerObs} restricting the sample
to stars with $\sigma_{\rm L}/L <0.2$. The slope is increased, but has a larger error bar ($3.05 \pm 0.19$) and
the rms is reduced to 0.44~mag. 

On the other hand, although the $T_{\rm eff}$ determinations based on the LDR method are precise (as discussed earlier),
possible systematic effects (which would also affect the determination of $\log g$ and $\log g_{\rm F}$)
could play a role \citep{Mancino20}.
For the subsample of 52 stars in L18 with five or more spectra, the cycle averaged $T_{\rm eff}$s (as quoted in
Table~\ref{Tab:FGLR}) are compared to the photometrically derived $T_{\rm eff}$s based on the SED fitting in G20.
The errors on the photometrically derived effective temperatures (the median is 180~K) are larger than those derived from spectroscopy.
There are two outliers Y Oph and S Vul, where the photometrically derived temperatures are considerably lower than those
quoted in L18 (570 and 830~K; $>4.3~\sigma$). For the other stars the difference
(spectroscopically - photometrically derived $T_{\rm eff}$) is $140 \pm 150$~K.

Systematic errors on the determination of the gravity could also play a role. The methodology used by L18 to determine
the stellar parameters, in particular $v_{\rm t}$ and gravity, is the standard one. A non-standard method is sometimes
also used in the literature, as introduced by \citet{KA99}. To avoid non-LTE-sensitive stronger Fe {\sc i} lines, $v_{\rm t}$ is derived
from Fe {\sc ii} lines and weak  Fe {\sc i} lines alone. This leads to higher $v_{\rm t}$, which in turn leads to higher gravities
when the ionisation balance is enforced.
For $\delta$ Cep \citet{KA99} find that the gravities are higher by 0.5~dex using the non-standard method.
The matter is also debated in \cite{Yong06}. They note that the non-standard method `has merits', but show that their derived gravities using
the standard method are self-consistent, one argument being that this gravity also produces ionisation equilibrium for Ti {\sc i} lines
that are more susceptible to non-LTE effects than Fe.
The non-standard method is also used in \cite{Takeda13}. \cite{Anderson16} excluded the gravities from that paper as they differed
from other sources they used.
Twelve stars overlap with the sample of stars with multi-epoch data from L18 (in Table~1).
\cite{Takeda13} present stellar parameters at between 7 and 17 epochs. The mean effective temperatures and mean gravities are calculated,
as well are FWGs at these epochs based on the data in \cite{Takeda13}, and fitted with low-order harmonic sine curves, as before,
to give the mean FWG.
The difference (min - max (mean)) between the parameters from the non-standard method minus those from the standard method are
$-8     - +444 \;(167)$~K in $T_{\rm eff}$, 
$+0.22 - +0.72 \;(+0.36)$~dex in $\log g$, and
$+0.11 - +0.67 \;(+0.34)$~dex in FWG,
with tendencies that the difference in all three quantities decreases with increasing period.

The FWGLR has the potential to be an alternative to the $PL$ relation in distance determination \citep{Anderson16}.
In its current empirically best calibrated version it is not. The scatter of 0.54~mag is larger than the
0.40~mag in the bolometric $PL$ relation determined in G20 using the identical sample of stars,
distances, and luminosities.

One issue is that the independent variable period is known with great precision, while the independent variable
FWG has a non-negligible error associated with it. The fitting of the FWG versus pulsation phase did not
provide more precise mean FWGs than simple averaging.
As the slope of the FWGLR is reasonably steep, any uncertainty on the FWG leads to a three times
larger uncertainty in $M_{\rm bol}$.

The discussion above also demonstrates that the stellar parameters should be derived in a uniform way.
  To exclude the influence of data analysis inhomogeneity altogether, Eq.~7 was re-determined using data only from L18.
  The usable sample is reduced to 161 stars and the slope and offset change marginally, less than 1$\sigma$.
  The standard approach used by L18 seems to give consistent results when considering the comparison to theory and the
  independent calibration of the FWGLR by \cite{K20}.
Changes in the FWG by $\sim +0.3-0.5$~dex, as implied by the non-standard method, would result in a disagreement. 

This paper is written with the tremendous potential offered by {\it Gaia} in mind.
Future data releases will provide information that will impact and improve on the results obtained here.
Primarily, improved parallaxes, taking into account binarity in the astrometrical solution, will provide
more precise distances and thus bolometric luminosities (e.g. through the SED fitting performed in G20).

Secondly, {\it Gaia} RVS spectra and {\it Gaia} Bp/Rp spectro-photometry will provide estimates of the stellar
parameters ($\log g$, $T_{\rm eff}$, also metallicity) in future releases. Only mean spectra in data release 3, and
epoch spectra in data release 4 \citep{Brown19}. An older analysis by \cite{Recio-Blanco16} indicate that end-of-mission
accuracies in $\log g$ of 0.1~dex or better can be reached in intermediate-metallicity F and G giants of
magnitude $G  \sim 10.3-11.8$ or brighter. Spectro-photometry can go fainter but with poorer accuracies
(0.2-0.4~dex in $\log g$ down to $G= 19$; Table~4 in \citealt{Bailer-Jones13}). As the nominal mission of 5 years
is extended, by +18 months until the end of 2020, and likely until the end of 2022,  these numbers should improve. 
In conclusion, the FWGLR could prove to become an extremely useful tool in Cepheid studies.

\begin{acknowledgements}
I would like to thank Dr. Bertrand Lemasle for interesting discussion on the determination of $\log g$ and
  commenting on a draft version of this paper.
This research has made use of the SIMBAD database and the VizieR catalogue access tool 
operated at CDS, Strasbourg, France.
\end{acknowledgements}

\bibliographystyle{aa.bst}
\bibliography{references.bib}

\begin{appendix}

\section{Additional material}
\label{App}

Additional fits for the FWGLR based on the models of \cite{Anderson16} are given in Table~\ref{Tab:fits}
for the three different metallicities, and with the slope fixed to the value in Eq.~\ref{Eq:And}.
The results for $Z= 0.006$ and $0.014$ agree within the error and justify the use of a single relation
combining the two metallicities (Eq.~\ref{Eq:And}). The $Z= 0.002$ models differ by a larger amount, qualitatively in
agreement with the remark in \cite{K20} on the fact that low metallicities (below $-0.6$~dex) have an effect
on the FWGLR.

\begin{table}[h]
  \centering

\caption{Fits of the type $M_{\rm bol} =  a \cdot (\log g - g_{\rm F}^{\rm b})  + b$.}
\begin{tabular}{ccrrl} \hline \hline 
     $a$          &        $b$        & rms & metallicity \\
\hline 
$3.381 \pm 0.025$ & $-3.031 \pm 0.014$ & 0.13 & $Z=0.014$ \\ 
$3.35$ fixed      & $-3.040 \pm 0.014$ & 0.14 & \\
$3.331 \pm 0.029$ & $-2.912 \pm 0.017$ & 0.16 & $Z=0.006$ \\ 
$3.35$  fixed     & $-2.905 \pm 0.014$ & 0.16 & \\
$3.426 \pm 0.026$ & $-2.732 \pm 0.016$ & 0.17 & $Z=0.002$ \\ 
$3.35$ fixed      & $-2.759 \pm 0.012$ & 0.18 & \\
\hline
\end{tabular} 
\label{Tab:fits}
\end{table}

\begin{figure}
  \centering

\begin{minipage}{0.49\textwidth}
\resizebox{\hsize}{!}{\includegraphics{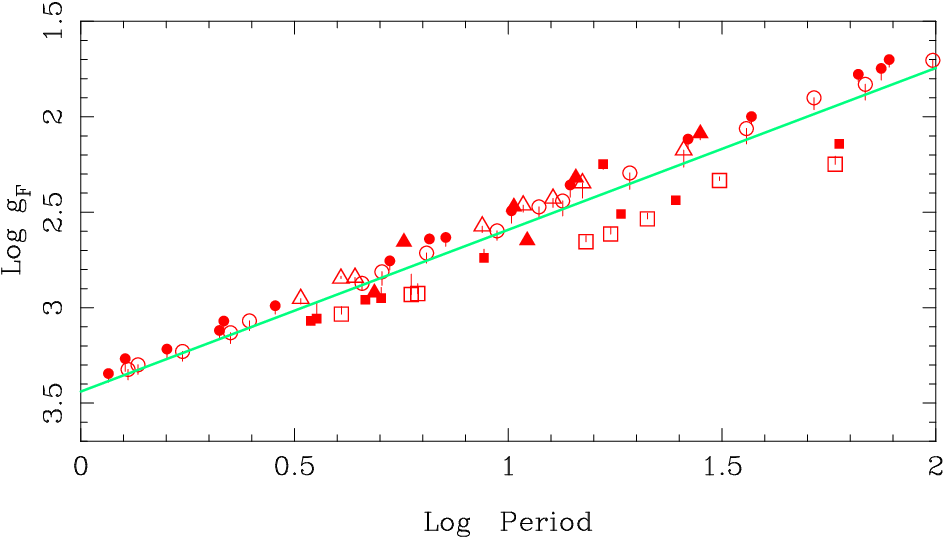}}
\resizebox{\hsize}{!}{\includegraphics{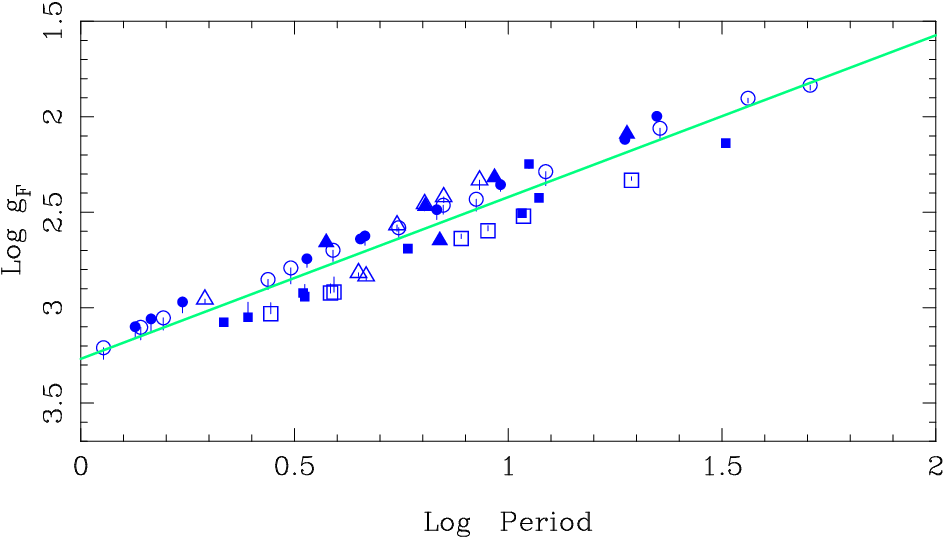}}
\end{minipage}

\caption{ 
  Relation between FWG and period for FU (top panel) and FO (bottom panel) models.
  The meaning of the symbols and colours is explained in Fig~\ref{Fig:FGWLR}.
  The green lines refer to the best fits, Eqs.~\ref{Eq:FWGPerFU} and \ref{Eq:FWGPerFO}.
  }
\label{Fig:FGWLR-FOU}
\end{figure}

\bigskip
\noindent
The bottom panel of Fig.~1 and Eq.~4 present the relation between FWG and pulsation period based on the models of \cite{Anderson16} with the
overtone periods converted to FU periods.
Figure~\ref{Fig:FGWLR-FOU} and Eqs.~\ref{Eq:FWGPerFU} and \ref{Eq:FWGPerFO} give the results for FU and FO pulsators separately.
The best fits are 
\begin{equation}
  \log g_{\rm F} =  (-0.847 \pm 0.015) \;\log P + (3.431  \pm 0.016)
\label{Eq:FWGPerFU}
\end{equation}
with an rms of 0.10~dex for the FU models, and 
\begin{equation}
  \log g_{\rm F} =  (-0.840 \pm 0.016) \;\log P + (3.255  \pm 0.013)
\label{Eq:FWGPerFO}
\end{equation}
with an rms of 0.08~dex for the FO models. 

\begin{figure}
  \centering
  
\begin{minipage}{0.33\textwidth}
\resizebox{\hsize}{!}{\includegraphics{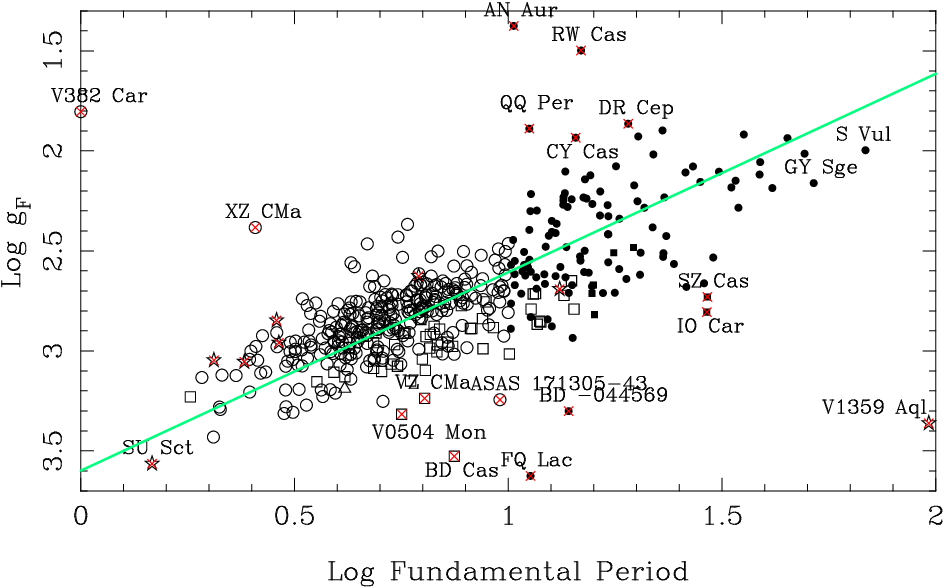}}
\resizebox{\hsize}{!}{\includegraphics{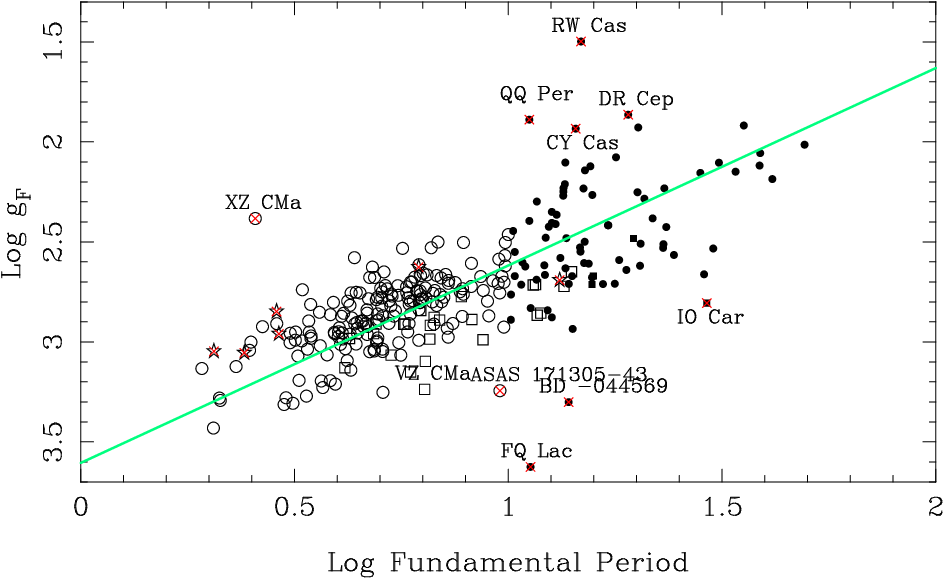}}
\resizebox{\hsize}{!}{\includegraphics{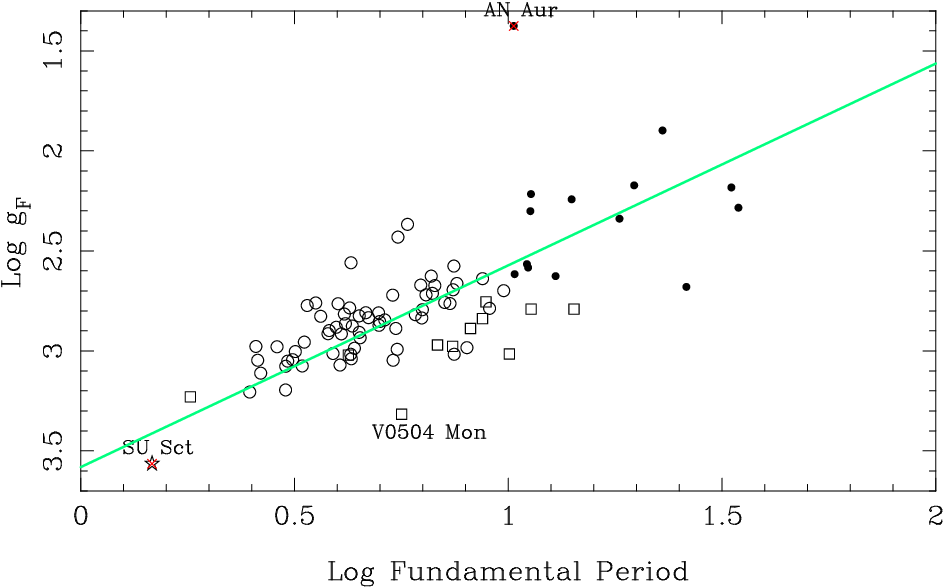}}
\resizebox{\hsize}{!}{\includegraphics{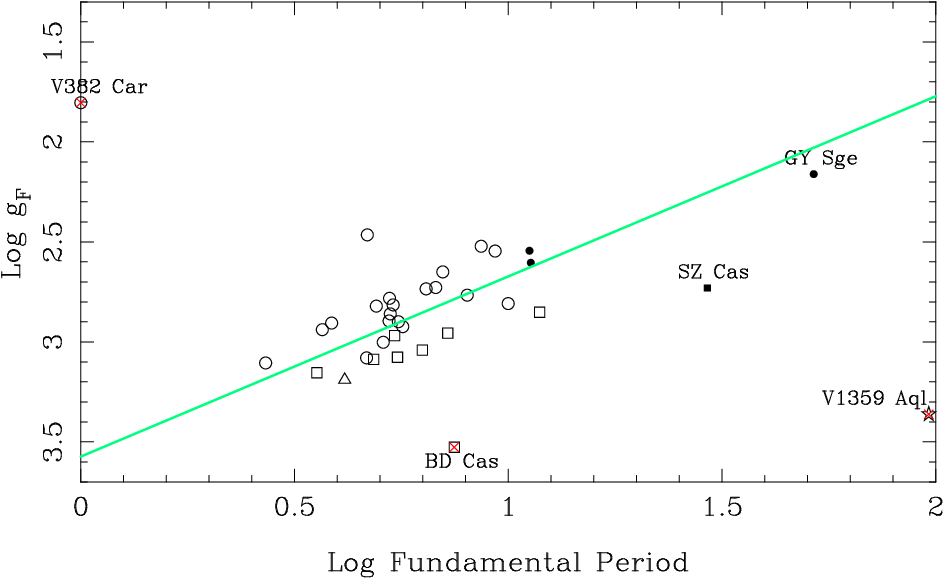}}
\resizebox{\hsize}{!}{\includegraphics{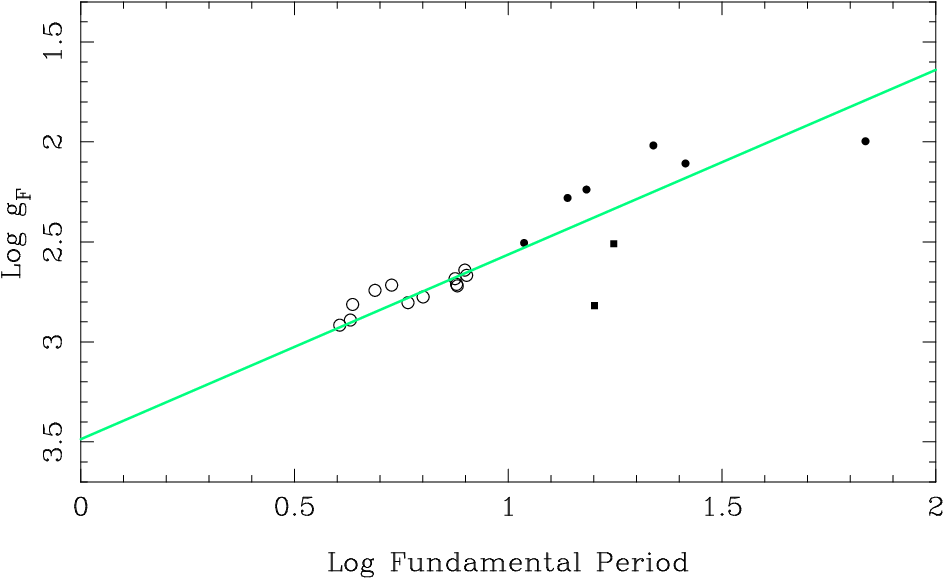}}
\resizebox{\hsize}{!}{\includegraphics{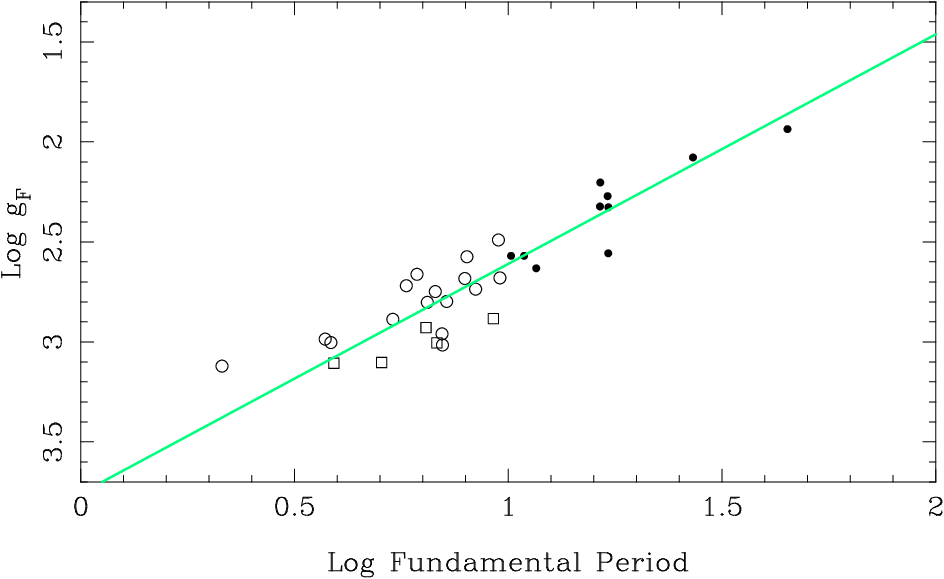}}
\end{minipage}

\caption{FWG vs period.
  The data from \cite{Genovali2014} is excluded in all plots.
  Different panels show different selections on the number of available $g_{\rm F}$ values.
  From top to bottom: all, $N_{\rm sp}=1$, $N_{\rm sp}=2$, $N_{\rm sp}=3-5$, $N_{\rm sp}=6-10$, $N_{\rm sp}\ge11$.
  The green lines refer to the best fits (see Table~\ref{Tab:FWGP}).
}
\label{App:Fig}
\end{figure}

\bigskip
\noindent
Additional fits for the relations between FWG and period are given in Table~\ref{Tab:FWGP} and are illustrated in Fig.~\ref{App:Fig}.
They show that when multiple $g_{\rm F}$ values are available the scatter in the relation decreases.
Assuming that the intrinsic scatter in the relation is 0.093~dex (Eq.~\ref{Eq:FWGPer}) a single determination
has an estimated error of about 0.13~dex (dominated by the error on $\log g$), while averaging six or more spectra
leads to an error of about 0.09~dex.

As noted in the main text, and illustrated by comparing Fig.~\ref{Fig:loggfper} and the top panel in Fig.~\ref{App:Fig}, a fair
fraction of outliers are stars with $T_{\rm eff}$ and $\log g$ taken from \cite{Genovali2014}.
\cite{Genovali2014} also present multiple observations for some stars, and XX Sgr and WZ Sgr are in common with
the subsample of stars in L18 with five or more available spectra. A comparison shows that the difference in $\log g_{\rm F}$
is dominated by the difference in $\log g$, that are of the order 0.5~dex. For some of the stars in the present sample the 
$\log g_{\rm F}$ (and $\log g$) values are too low by 1 dex.
As they seem to use the same methodology as L18 in deriving the stellar parameters, no simple explanation is offered to
explain this discrepancy.

\begin{table*}[h]
  \centering

\caption{Fits of the type $g_{\rm F} =  a \cdot \log P  + b$.}
\begin{tabular}{ccrrl} \hline \hline 
     $a$          &        $b$        & rms & N & Remarks \\
\hline 
$-0.802 \pm 0.028$ & $3.436 \pm 0.025$ & 0.159 & 443 & standard, Eq.~\ref{Eq:FWGPerObs} \\ 
$-0.804 \pm 0.028$ & $3.438 \pm 0.025$ & 0.158 & 442 & excluding \cite{Genovali2014} \\
$-0.793 \pm 0.036$ & $3.438 \pm 0.032$ & 0.160 & 275 & $N_{\rm sp}=1$, excluding \cite{Genovali2014} \\
$-0.752 \pm 0.073$ & $3.381 \pm 0.060$ & 0.172 &  87 & $N_{\rm sp}=2$, excluding \cite{Genovali2014} \\
$-0.630 \pm 0.116$ & $3.351 \pm 0.100$ & 0.160 &  32 & $N_{\rm sp}=3-5$, excluding \cite{Genovali2014} \\
$-0.805 \pm 0.106$ & $3.370 \pm 0.110$ & 0.139 &  20 & $N_{\rm sp}=6-10$, excluding \cite{Genovali2014} \\
$-1.046 \pm 0.088$ & $3.663 \pm 0.086$ & 0.125 &  31 & $N_{\rm sp}\ge11$, excluding \cite{Genovali2014} \\
$-0.970 \pm 0.063$ & $3.560 \pm 0.064$ & 0.127 &  50 & $N_{\rm sp}\ge6$, excluding \cite{Genovali2014} \\
\hline
\end{tabular} 
\label{Tab:FWGP}
\end{table*}

\bigskip
\noindent
Table~\ref{Tab:Sam} compiles the FWG and luminosity data for the entire sample.
The full table is available at the CDS.

\begin{table*}[b]
  \centering

\setlength{\tabcolsep}{1.6mm}
\caption{FWG data for the entire sample (first entries only).}
\begin{tabular}{rlrccrrccccccr} \hline \hline 
  Name             & Type     & Period &  $d$   & $\sigma_{\rm d}$ &    $L$    & $\sigma_{\rm L}$ & $N_{\rm spec}$ & $\log g_{\rm F}$ & $\sigma_{\rm \log g_{\rm F}}$ & Min-Max & Ref.  \\
                   &          & (days) &  (kpc) &    (kpc)        &  (\lsol) &     (\lsol)      &              &     (cgs)        &   (cgs)                   &   (cgs) &  \\
  \hline 
   AA Gem          & DCEP     &  11.302 &  3.400 &  0.829 &   3400.0 &    122.7 &  2 & 2.216 & 0.11 & 0.04 & 1        &  \\ 
   AA Mon          & DCEP     &   3.938 &  3.922 &  0.709 &    922.8 &     33.6 &  1 & 3.211 & 0.16 & -   & 1        &  \\ 
   AB Cam          & DCEP     &   5.788 &  4.200 &  0.966 &   1463.5 &     79.3 &  1 & 2.754 & 0.15 & -   & 1        &  \\ 
   AC Mon          & DCEP     &   8.014 &  2.400 &  0.400 &   1991.6 &     42.2 &  4 & 2.766 & 0.08 & 0.21 & 1        &  \\ 
   AD Cam          & DCEP     &  11.261 &  4.600 &  0.756 &   2048.8 &     87.0 &  2 & 2.301 & 0.11 & 0.07 & 1        &  \\ 
   AD Cru          & DCEP     &   6.398 &  2.994 &  0.394 &   1881.9 &     93.2 &  1 & 2.730 & 0.15 & -   & 1        &  \\ 
   AD Gem          & DCEP     &   3.788 &  2.500 &  0.673 &    966.0 &     32.0 &  2 & 2.914 & 0.11 & 0.12 & 1        &  \\ 
   AD Pup          & DCEP     &  13.596 &  4.100 &  0.946 &   4650.8 &    356.8 &  1 & 2.103 & 0.15 & -   & 1        &  \\ 
   AE Tau          & DCEP     &   3.897 &  3.367 &  0.606 &    953.2 &     11.9 &  1 & 2.802 & 0.15 & -   & 1        &  \\ 
   AE Vel          & DCEP     &   7.134 &  2.100 &  0.187 &   1842.6 &    169.2 &  1 & 2.663 & 0.15 & -   & 1        &  \\ 
   AG Cru          & DCEP     &   3.837 &  1.506 &  0.094 &   1773.5 &     49.7 &  1 & 2.864 & 0.15 & -   & 1        &  \\ 
   AH Vel          & DCEPS    &   4.227 &  0.752 &  0.035 &   2604.0 &     37.7 &  2 & 2.838 & 0.11 & 0.04 & 1        &  \\ 
   alpha UMi       & DCEPS    &   3.970 &  0.133 &  0.002 &   2410.9 &    622.8 &  2 & 2.888 & 0.11 & 0.14 & 2,3      &  \\ 
   AN Aur          & DCEP     &  10.291 &  3.400 &  0.985 &   3124.5 &     58.2 &  2 & 2.630 & 0.11 & 0.24 & 1        &  \\ 
   AO Aur          & DCEP     &   6.763 &  3.400 &  0.839 &   1620.9 &     49.3 &  3 & 2.728 & 0.09 & 0.18 & 1        &  \\ 
   AO CMa          & DCEP     &   5.816 &  3.600 &  0.434 &   1197.9 &     58.1 &  1 & 2.950 & 0.16 & -   & 1        &  \\ 
   AP Pup          & DCEP     &   5.084 &  1.183 &  0.051 &   2579.5 &     87.4 &  1 & 2.869 & 0.15 & -   & 1        &  \\ 
   AP Sgr          & DCEP     &   5.058 &  0.861 &  0.041 &   1651.8 &     38.7 &  1 & 2.780 & 0.15 & -   & 1        &  \\ 
   AQ Car          & DCEP     &   9.769 &  3.030 &  0.317 &   3837.4 &    289.0 &  1 & 2.702 & 0.15 & -   & 1        &  \\ 
   AQ Pup          & DCEP     &  30.149 &  2.900 &  0.336 &  11481.5 &    330.8 &  1 & 2.533 & 0.15 & -   & 1        &  \\ 
   AS Per          & DCEP     &   4.973 &  1.200 &  0.087 &   1042.0 &     36.7 &  2 & 2.810 & 0.11 & 0.30 & 1        &  \\ 
   AT Pup          & DCEP     &   6.665 &  1.637 &  0.085 &   2495.6 &    194.9 &  1 & 2.757 & 0.15 & -   & 1        &  \\ 
   AV Cir          & DCEPS    &   3.065 &  0.944 &  0.033 &   2169.7 &     85.7 &  1 & 2.843 & 0.15 & -   & 1        &  \\ 
   AV Sgr          & DCEP     &  15.415 &  2.100 &  0.287 &   4413.1 &    139.5 &  1 & 2.609 & 0.15 & -   & 1        &  \\ 
   AW Per          & DCEP     &   6.464 &  0.700 &  0.044 &   1646.8 &     76.6 & 11 & 2.802 & 0.05 & 0.47 & 1        &  \\ 
   AX Cir          & DCEP     &   5.273 &  0.500 &  0.151 &   1854.6 &     33.1 &  3 & 2.782 & 0.09 & 0.08 & 1        &  \\ 
   AX Vel          & DCEP(B)  &   2.593 &  1.517 &  0.077 &   1750.2 &    166.6 &  2 & 3.047 & 0.11 & 0.05 & 1        &  \\ 
   AY Cen          & DCEP     &   5.310 &  1.689 &  0.100 &   1864.4 &    303.0 &  1 & 2.821 & 0.15 & -   & 1        &  \\ 
   AZ Cen          & DCEPS    &   3.212 &  2.137 &  0.158 &   2017.4 &     50.1 &  1 & 2.986 & 0.15 & -   & 1        &  \\ 
   BB Cen          & DCEPS    &   3.998 &  3.610 &  0.363 &   3100.8 &    110.7 &  1 & 2.888 & 0.15 & -   & 1        &  \\ 
   BB Gem          & DCEP     &   2.308 &  4.082 &  0.825 &   1135.9 &     49.5 &  1 & 3.123 & 0.16 & -   & 1        &  \\ 
   BB Her          & DCEP     &   7.508 &  3.623 &  0.759 &   3122.0 &    153.2 &  8 & 2.684 & 0.05 & 0.32 & 1        &  \\ 
   BB Sgr          & DCEP     &   6.637 &  0.700 &  0.023 &   1529.1 &     30.8 &  1 & 2.800 & 0.15 & -   & 1        &  \\ 
   BC Pup          & DCEP     &   3.544 &  6.500 &  1.109 &    938.2 &     64.4 &  2 & 2.760 & 0.11 & 0.13 & 4,(17)   &  \\ 
\hline
\end{tabular} 
\tablefoot{
  Column~1: Name.
  Column~2: Type (from Table~1 in G20). Nomenclature follows that used by the
    VSX \citep{Watson06}\footnote{described in \url{https://www.aavso.org/vsx/index.php?view=about.vartypes}.} .
  Column~3: Period (from G20).
  Column~4: Distance (from G20).
  Column~5: Error on distance (from G20).
  Column~6: Luminosity (from G20).
  Column~7: Error on Luminosity (from G20).
           The error is the fit error, and does not include the error on the distance.
           If the total error on $L$ is desired it can be calculated from $\sqrt{\sigma_{\rm L}^2 + \Delta^2}$ with
           $\Delta = L \cdot ((1 + \sigma_{\rm d}/d)^2 -1)$. The total error is plotted in Fig.~\ref{Fig:loggfobs}.
  Column~8: Number of available spectra, $N_{\rm spec}$.
  Column~9: Average of available $\log g_{\rm F}$ values.
  Column~10: Estimated error on the average $\log g_{\rm F}$ value.
    This includes the error on $T_{\rm eff}$ (when not given in the reference a conservative value of 100~K has been used) and the error
    on $\log g$ (assumed to be 0.15~dex, unless given specifically), divided by $\sqrt{N_{\rm spec}}$.
  Column~11: Difference between highest and lowest $\log g_{\rm F}$ value. 
  Column~12: References for $\log g$ and $T_{\rm eff}$ values to calculate $\log g_{\rm F}$ and error: 
  (1) \cite{Luck18}, (2) \cite{And1994}, (3) \cite{Boyarchuk81}, (4) \cite{Luck03}, (5) \cite{Schmidt11}, (6) \cite{And02III},
  (7) \cite{Lemasle08}, (8) \cite{And13}, (9) \cite{Kov05}, (10) \cite{Luck06}, (11) \cite{Yong06}, (12) \cite{Lemasle15},
  (13) \cite{Romaniello08}, (14) \cite{And04}, (15) \cite{Lemasle07}, (16) \cite{And02I}, (17) \cite{Genovali2014},
  (18) \cite{starhorse}, (19) \cite{MA15}, (20) \cite{Andrievsky2016}, (21) \cite{Inno19}.
 Numbers in parentheses indicate references not considered. 
}
\label{Tab:Sam}
\end{table*}

\end{appendix}

\end{document}